\documentclass[conference]{IEEEtran}
\usepackage{graphicx}
\usepackage{placeins}
\usepackage{hyperref}
\usepackage{amsmath}
\usepackage{pifont}
\usepackage{float}
\usepackage{llncsconf}
\usepackage{array}
\usepackage{multirow}

\usepackage{multicol}
\usepackage[per-mode=symbol]{siunitx}
\usepackage{comment}
\usepackage{siunitx}
\usepackage{caption}
\usepackage{subcaption}

\usepackage{graphicx}

\usepackage{siunitx}
\usepackage{url}

\usepackage{cite}
\usepackage{amsmath,amssymb,amsfonts}
\usepackage{algorithmic}
\usepackage{textcomp}
\usepackage{xcolor}

\def\BibTeX{{\rm B\kern-.05em{\sc i\kern-.025em b}\kern-.08em
    T\kern-.1667em\lower.7ex\hbox{E}\kern-.125emX}}

\makeatletter
\newcommand{\newlineauthors}{%
  \end{@IEEEauthorhalign}\hfill\mbox{}\par
  \mbox{}\hfill\begin{@IEEEauthorhalign}
}

\usepackage{tikz}

\newcommand\copyrighttext{%
  \footnotesize \textcopyright \the\year{} IEEE. Personal use of this material is permitted. Permission from IEEE must be obtained for all other uses, including reprinting/republishing this material for advertising or promotional purposes, collecting new collected works for resale or redistribution to servers or lists, or reuse of any copyrighted component of this work in other works.}

\newcommand\copyrightnotice{%
\begin{tikzpicture}[remember picture,overlay]
\node[anchor=south,yshift=10pt] at (current page.south) {\fbox{\parbox{\dimexpr0.75\textwidth-\fboxsep-\fboxrule\relax}{\copyrighttext}}};
\end{tikzpicture}%
}

\def\BibTeX{{\rm B\kern-.05em{\sc i\kern-.025em b}\kern-.08em
    T\kern-.1667em\lower.7ex\hbox{E}\kern-.125emX}}
\begin{document}
\title{Enhancing Traffic Safety with AI and 6G: Latency Requirements and Real-Time Threat Detection}
\thanks{Identify applicable funding agency here. If none, delete this.}

\author{\IEEEauthorblockN{1\textsuperscript{st} Kurt Horvath}
\IEEEauthorblockA{\textit{ITEC} \\
\textit{University of Klagenfurt}\\
Klagenfurt, Austria \\
0009-0008-7737-7013 \\
kurt.horvath@aau.at}
~\\
\and
\IEEEauthorblockN{2\textsuperscript{nd} Dragi Kimovski}
\IEEEauthorblockA{\textit{ITEC} \\
\textit{University of Klagenfurt}\\
Klagenfurt, Austria \\
0000-0001-5933-3246 \\
dragi.kimovski@aau.at }
~\\
\and
\IEEEauthorblockN{3\textsuperscript{nd} Stojan Kitanov}
\IEEEauthorblockA{\textit{Faculty of Information Sciences} \\
\textit{Mother Teresa University}\\
Skopje, R. N. Macedonia \\
0000-0002-7222-1078}
~\\
\and
\IEEEauthorblockN{4\textsuperscript{rd} Radu Prodan}
\IEEEauthorblockA{\textit{Department of Computer Science} \\
\textit{University of Innsbruck}\\
Innsbruck, Austria \\
0000-0002-8247-5426}

}

\maketitle
\copyrightnotice

\begin{abstract}
The rapid digitalization of urban infrastructure opens the path to smart cities, where IoT-enabled infrastructure enhances public safety and efficiency. This paper presents a 6G and AI-enabled framework for traffic safety enhancement, focusing on real-time detection and classification of emergency vehicles and leveraging 6G as the latest global communication standard. The system integrates sensor data acquisition, convolutional neural network-based threat detection, and user alert dissemination through various software modules of the use case. We define the latency requirements for such a system, segmenting the end-to-end latency into computational and networking components. Our empirical evaluation demonstrates the impact of vehicle speed and user trajectory on system reliability. The results provide insights for network operators and smart city service providers, emphasizing the critical role of low-latency communication and how networks can enable relevant services for traffic safety.

\end{abstract}

\begin{IEEEkeywords}
6G, traffic safety, network latency, artificial intelligence
\end{IEEEkeywords}


\section{Introduction}\label{sec:Introduction}
The digitalization of urban areas marks a critical milestone in the evolution of technologically driven smart cities. Smart cities, as a prominent domain within the Internet of Things (IoT), enable seamless collaboration between technology providers and public authorities to establish, monitor, and manage extensive networks of sensors and actuators embedded within urban infrastructure.

Sharing sensor data across these networks offers an opportunity to create various services, notably improving the visual range for vehicles and drivers. Advanced wireless communication technologies \cite{michailidou2022equality}, such as 5G, 6G, and WiFi7, facilitate data transmission from stationary sensor modules to user module devices or vehicle control systems\cite{kocher13}. Among these, 6G stands out for its potential to minimize latency, making it a pivotal enabler for such applications. This work leverages 6G to connect key system components, including the \textbf{Sensor Module}, responsible for data acquisition, preprocessing, and transmission, and the \textbf{Processing Module}, which employs a convolutional neural network (CNN)-specifically Darknet with YOLOv4 to detect emergency vehicles. Upon detection, the system triggers an event, categorizes it based on thread classification, and alerts users through the \textbf{Consumer Module}, which interacts with them visually, acoustically, or via direct vehicle control (e.g., braking).
This paper characterizes the network latency of 6G networks to enable this traffic safety use case. First, we define the end-to-end latency ($T_{tot}$) required to operate the system effectively. Next, we isolate the networking latency by accounting for computational delays, leaving the permissible network latency. Finally, we allocate this latency across logical networking steps within the workflow defined by the use case.

Our implementation analyzes how factors such as vehicular velocity and user trajectory affect the system's reliability and the accuracy of alerts. We present a model to detect and classify threats based on user and sensor positions in their vicinity. The main contributions of this work are:
\begin{itemize}
    \item Definition of latency requirements for 6G communication networks in a real-world traffic safety use case.
    \item Design of a framework that integrates AI and 6G to enhance traffic safety through advanced threat classification.
    \item Empirical evaluation of the use case to derive actionable parameters for network operators and smart city service providers.
\end{itemize}

The paper organizes its content as follows: Section \ref{sec:relatedWork} explores the definitions of real-time requirements in traffic safety systems. Section \ref{sec:Model} outlines the temporal components of the proposed framework and the detection of intersections and proximities between sensors and users. Section \ref{sec:Scenario} elaborates on threat classification in the traffic safety use case. Section \ref{sub:requirements} discusses latency contributions of software modules and their implications for 6G network requirements. Section \ref{sec:framework} details the implementation workflow. Finally, Section \ref{sec:Results} evaluates vehicle classification using AI and threat classification, while Section \ref{sec:Conclusion} defines tolerable network latency and its implications for existing network infrastructure.

\section{Related Work}\label{sec:relatedWork}
This section provides an overview of use cases where artificial intelligence (AI) enhances traffic safety and defines the term "real-time" for safety-relevant applications. We establish a uniform terminology to identify the time components addressed in the referenced works. This uniformity forms the basis for analyzing the identified value ranges and assessing their applicability.

Lujic \cite{lujic21} contributes \textit{InTraSafEd5G} (Increasing Traffic Safety with Edge and 5G), a system designed to generate proximate traffic warnings. The system evaluates its performance using AI models while employing 5G as the primary communication network for disseminating information. Lujic identifies latency as the key metric, categorizing it into communication latency $\{T_{eval}, T_{exe}, T_{act}\}$ and computing latency $\{T_S, T_P, T_C\}$. The evaluations primarily focus on 5G networks but include supplementary assessments using 3G and 4G networks. The results reveal that 4G and 5G achieve latencies below \SI{24}{\milli\second}, whereas 3G exceeds this threshold. The system uses the MobileNet SSD framework, a convolutional neural network (CNN), for detection, which introduces a processing latency.
\\
Xu \cite{xu2022real} proposes an alternative method operating within the 5G-V2X framework. This approach uses an onboard camera system to orient drivers on upcoming roadways. As a visual guidance system, it enhances traffic safety by providing directional alignment information to the driver. The method includes a preliminary data filtration stage, ensuring the artificial intelligence system processes only relevant data critical for road direction detection. Various AI frameworks handle the detection process, which is executed off-vehicle. Although the system does not directly control the vehicle, it achieves an average total latency $T_{tot}$ of \SI{150}{\milli\second}.
\\
Rausch \cite{rausch2021towards} presents a platform that provides supplementary information to enhance the user's perception of the road. This platform extends the user's field of view by rendering shapes of objects beyond their direct line of sight. The system comprises a producer module integrated into a camera system and a consumer module implemented via Augmented Reality (AR) equipment. AI processes occur at every stage of the computing continuum, which researchers have extensively evaluated. Instead of isolating the latency of individual components, the study measures the total latency $T_{tot}$. The observations indicated that accessing Microsoft Azure Cloud via 4G resulted in the highest mean latency values at \SI{1116}{\milli\second} while accessing Microsoft Azure Cloud through other means showed a mean latency of \SI{678}{\milli\second}. The lowest latency (\SI{146}{\milli\second}) was achieved using an edge device over WiFi. While these results provide a performance spectrum for the given real-world scenario, the study does not explicitly determine whether users perceived the performance as real-time.
\\
Wei \cite{Wei19} highlights the challenges of vehicle tracking and traffic flow estimation in smart cities, emphasizing the need for more efficient solutions given the constraints of existing systems. Many current approaches rely on high frame rates and high-resolution video, often unavailable in cities due to bandwidth and storage limitations. For example, New York City's traffic cameras operate at extremely low frame rates (e.g., one frame per second to 5 Hz) and low resolution (352x240), rendering traditional tracking methods ineffective. Wei introduced LFTSys (Low Frame-rate Tracking System), designed explicitly for low Frame-rate camera networks to overcome these limitations.

\paragraph{Real-Time Definition Based on Review}\label{par:realtimeDefinition}
The reviewed material reveals a wide range of definitions for real-time, which vary based on the sensor type and application. Xu \cite{xu2022real} and Wei \cite{Wei19} both define real-time in the range of 150 to \SI{200}{\milli\second} (5 Hz). Given the proximity of their work to the domain addressed in our implementation, we adopt the same definition for $T_{tot}$.

\section{Model}\label{sec:Model}
The model consists of two main sections. First, we identify the time components involved in the use case workflow. Next, we analyze how to determine whether a user will intersect with the vehicle's extended visual range, as provided by the sensor. We discuss the implications of proximity when vehicles intersect and, finally, describe how we detect the driving direction using a single frame from the sensor.

\subsection{Temporal Model}\label{sub:temporalModel}
\begin{figure}[t]
   \centering
    \includegraphics[width=0.4\textwidth]{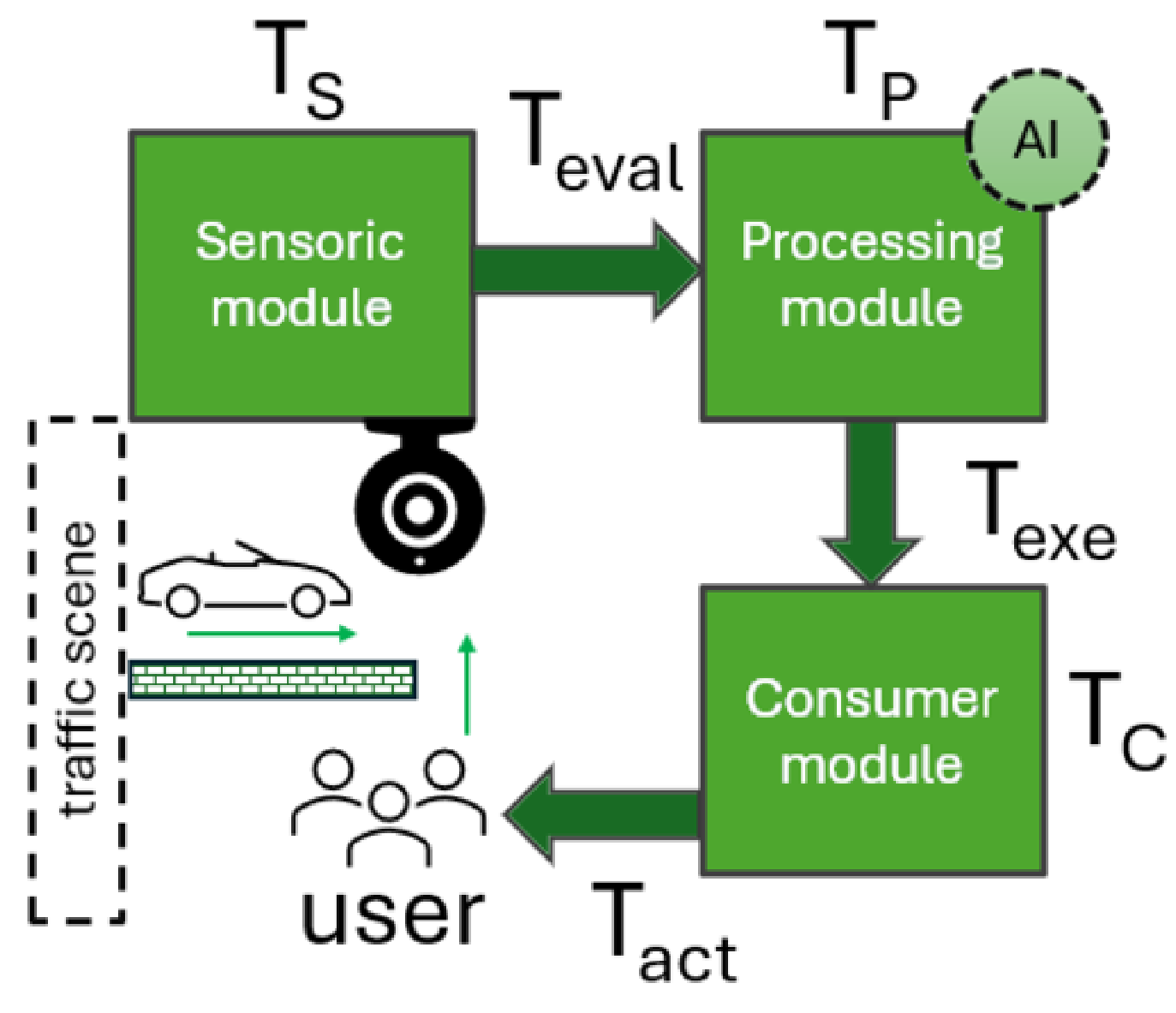} 
\caption{Topological overview of AI-based traffic control system.}
    \label{fig:geocontext}%
    \vspace{-8mm}
\end{figure}

We divide the process into specific time spans to evaluate how solutions support real-time traffic safety applications. We represent the total processing time $T_{tot}$, as illustrated in Figure \ref{fig:geocontext}, with the following equation:

\begin{equation} \label{eq:timecomponents}
T_{tot} = T_S + T_{eval} + T_P + T_{exe} + T_C + T_{act}. 
\end{equation} 

Where, $T_S$ represents the latency of the sensor during data acquisition.\\
$T_{eval}$ denotes the duration required to transfer data to the AI-processing module via the 6G network.\\
$T_P$ refers to the time needed for feature detection using a pre-trained model. The processing module also evaluates detected features and formulates decisions based thread classification. \\
$T_{exe}$ describes the time required to transmit the alarm event to the user module on the road.\\
$T_C$ is the time taken to validate the alarm.\\
$T_{act}$ represents the time until the requested action is performed. This delay can be significant if the action involves human intervention.

\paragraph{Real-Time Constraints}
To achieve real-time performance for the given scenario, we define the following constraint:

\begin{equation} \label{eq:realtime}
T_{min} \leq T_{rt} \leq T_{tot}.
\end{equation} 

Where $T_{min}$ signifies the theoretical minimal delay, and $T_{rt}$ represents the maximum permissible technical delay (e.g., signal propagation, processing overhead). This constraint ensures real-time processing for the use case.
\subsection{Spatial Model}\label{sub:spatialModel}

\subsubsection{Trajectories and Vectors}

The user's trajectory originates at point coordinates \((\phi_u, \lambda_u)\), which are converted into Cartesian coordinates \((x_u, y_u, z_u)\). The user's position vector at time $\vec{U}(t)$ is expressed as:

\begin{equation} \label{eq:usertraj}
\vec{U}(t) = \vec{P}_u + t 
\begin{bmatrix} 
\cos \theta_u \\ 
\sin \theta_u 
\end{bmatrix}
\end{equation}

where \(\vec{P}_u = \begin{bmatrix} x_u \\ y_u \end{bmatrix}\) represents the user's starting position and $\theta_u$ the angle of the vector in Cartesian coordinates. 

Similarly, the sensor's trajectory starts at coordinates \((\phi_s, \lambda_s)\), which are also converted to Cartesian coordinates \((x_s, y_s, z_s)\). The sensor's position vector at time $\vec{S}(t) $ is given by:
\vspace{-3mm}
\begin{equation} \label{eq:sensortraj}
\vec{S}(t) = \vec{P}_s + t 
\begin{bmatrix} 
\cos \theta_s \\ 
\sin \theta_s 
\end{bmatrix}
\end{equation}

where \(\vec{P}_s = \begin{bmatrix} x_s \\ y_s \end{bmatrix}\) denotes the sensor's starting position and $\theta_u$ the angle of the vector in Cartesian coordinates.

\subsubsection{Intersection Point}

The trajectories intersect if a point exists where the user and sensor vectors are equal:

\begin{equation} \label{eq:intersection1}
\vec{U}(t_u) = \vec{S}(t_s)
\end{equation}

Expanding this relation, we derive:

\begin{equation} \label{eq:intersection2}
\vec{P}_u + t_u 
\begin{bmatrix} 
\cos \theta_u \\ 
\sin \theta_u 
\end{bmatrix} = 
\vec{P}_s + t_s 
\begin{bmatrix} 
\cos \theta_s \\ 
\sin \theta_s 
\end{bmatrix}
\end{equation}

This leads to the following system of equations:

\begin{equation} \label{eq:intersection3}
\begin{aligned}
  x_u + t_u \cos \theta_u &= x_s + t_s \cos \theta_s \\
  y_u + t_u \sin \theta_u &= y_s + t_s \sin \theta_s
\end{aligned}
\end{equation}

We represent this system in matrix form:

\begin{equation} \label{eq:intersection4}
\begin{bmatrix}
\cos \theta_u & -\cos \theta_s \\
\sin \theta_u & -\sin \theta_s
\end{bmatrix}
\begin{bmatrix}
 t_u \\
 t_s
\end{bmatrix} =
\begin{bmatrix}
 x_s - x_u \\
 y_s - y_u
\end{bmatrix}
\end{equation}

Solving for \(t_u\) and \(t_s\) provides the times when the trajectories intersect but not yet the precise location of the intersection.

We define the point of intersection as \(I\), with components \(x_i, y_i\) in the Cartesian coordinate system:

\begin{equation} \label{eq:intersectionPoint}
I = \begin{bmatrix}
x_i \\
y_i
\end{bmatrix} = \vec{P}_u + t_u 
\begin{bmatrix} 
\cos \theta_u \\ 
\sin \theta_u 
\end{bmatrix}
\end{equation}

\subsection{Intra-Cell Intersection}\label{sub:intracell}

\begin{figure}[ht]
\vspace{-4mm}
  \centering
  \includegraphics[width=0.45\textwidth]{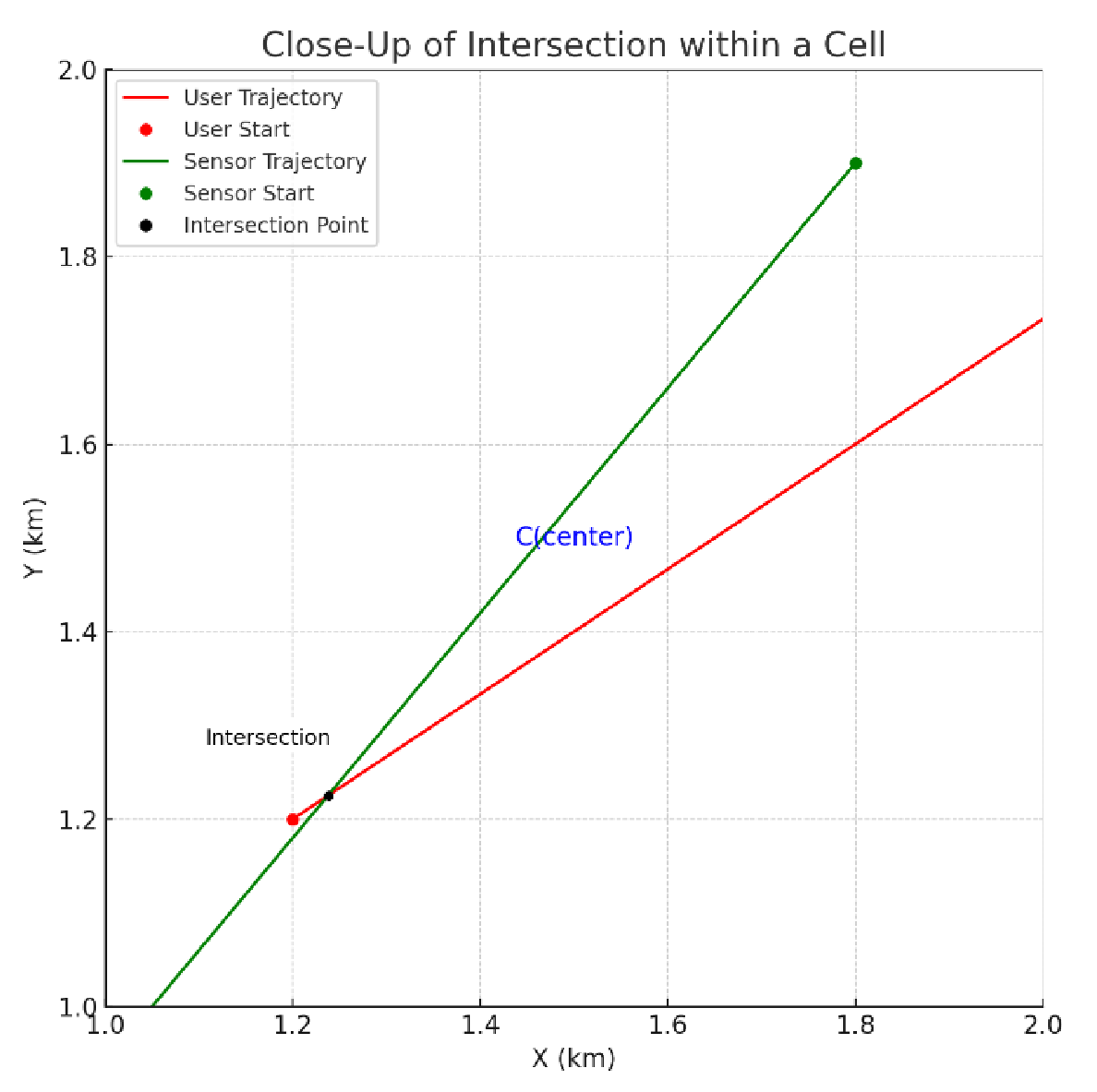}
  \vspace{-4mm}
  \caption{Illustration of an intra-cell intersection. The user and sensor trajectories intersect within the boundaries of the cell.}
  \label{fig:intersection}
  \vspace{-4mm}
\end{figure}

For an intra-cell intersection, as shown in Figure \ref{fig:intersection}, the user and sensor trajectories intersect within the same cell where the user resides. Let the user's cell \(C\) be defined by its bottom-left corner \((x_{\text{min}}, y_{\text{min}})\) and a size of \(d = 1\,\text{km}\). The cell expresses its bounds as:
\vspace{-3mm}
\begin{equation} \label{eq:intracell}
\text{intra} = \left(
\begin{aligned}
  x_{\text{min}} &\leq x_I \leq x_{\text{min}} + d, \\
  y_{\text{min}} &\leq y_I \leq y_{\text{min}} + d
\end{aligned}
\right)
\end{equation} 

\subsection{Inter-Cell Intersection}\label{sub:intercell}

\begin{figure}[ht]
\vspace{-4mm}
  \centering
  \includegraphics[width=0.45\textwidth]{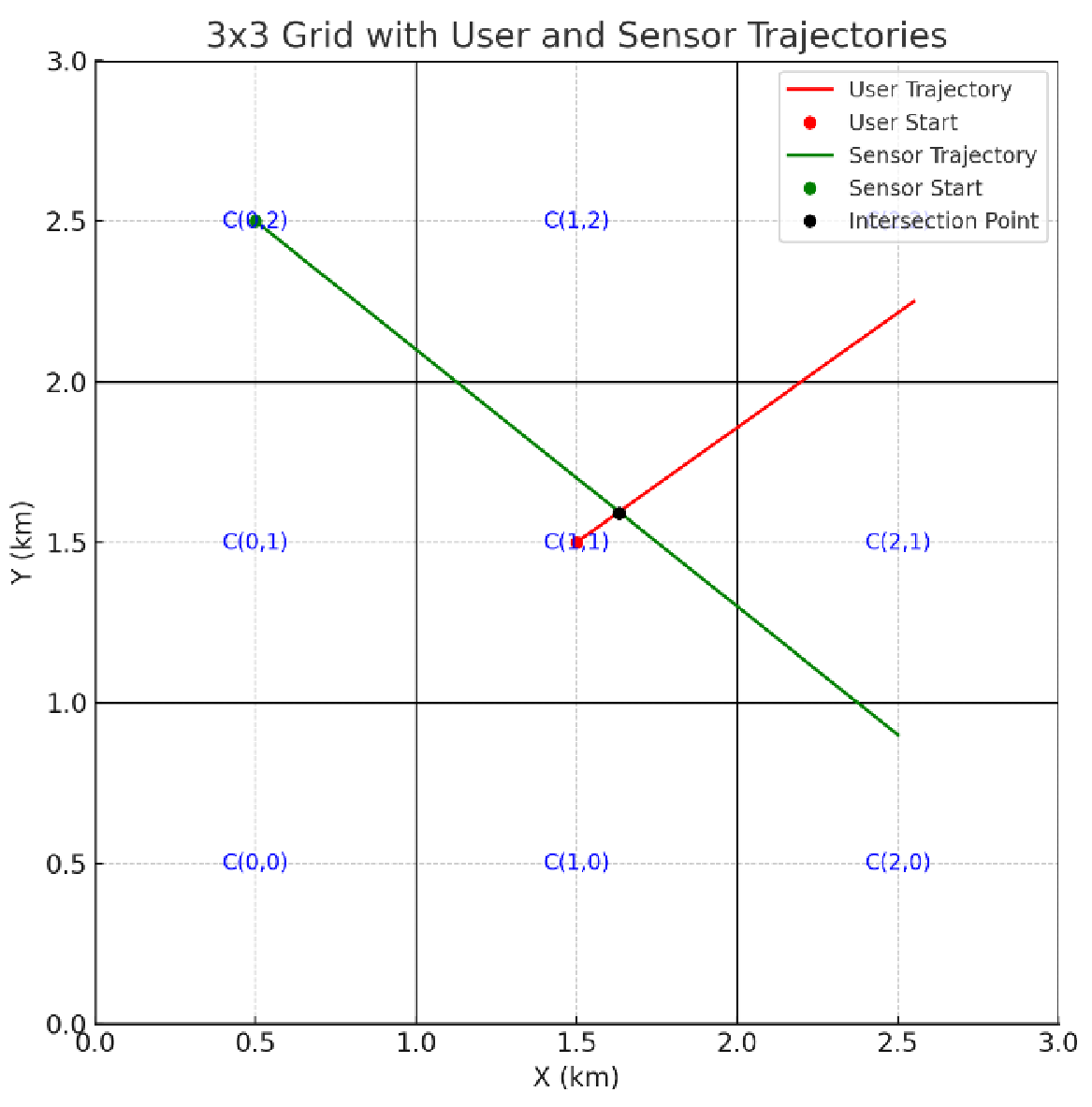}
  \vspace{-2mm}
  \caption{Example of a \(3 \times 3\) grid. The user's cell is at the centre while neighbouring cells represent potential inter-cell intersections.}
  \label{fig:grid}
  \vspace{-6mm}
\end{figure}

For an inter-cell intersection, the sensor's trajectory may originate from one of the neighbouring cells within a \(3 \times 3\) grid \(G\), as illustrated in Figure \ref{fig:grid} and defined in Equation \ref{eq:grid}. The grid positions the user's cell at the centre, defining neighbouring cells by offsets relative to the user's cell.
\begin{equation} \label{eq:grid}
G = \{C_{i+dx, j+dy} \mid dx, dy \in \{-1, 0, 1\}, (dx, dy) \neq (0, 0)\}
\end{equation} 

Each neighbouring cell has bounds similar to the user's but shifted according to the offsets. For instance, the cell at \((i+1, j-1)\) has bounds defined as:
\vspace{-3mm}
\begin{equation} \label{eq:interDef}
\text{inter} = \left(
\begin{aligned}
  x_{\text{min}} + d \cdot dx &\leq x_i \leq x_{\text{min}} + d \cdot (dx + 1), \\
  y_{\text{min}} + d \cdot dy &\leq y_i \leq y_{\text{min}} + d \cdot (dy + 1)
\end{aligned}
\right)
\end{equation}

\subsection{Intra-Cell Coexistence}\label{sub:coexistence}

Intra-cell coexistence requires the sensor and the user to reside within the same cell. Since the user knows their residence cell, only the sensor's location needs validation to confirm coexistence. The bounds for coexistence are:
\vspace{-1mm}
\begin{equation} \label{eq:coexist}
\text{coexist} = \left(
\begin{aligned}
  x_{\text{min}} + d \cdot dx &\leq x_s \leq x_{\text{min}} + d \cdot (dx + 1), \\
  y_{\text{min}} + d \cdot dy &\leq y_s \leq y_{\text{min}} + d \cdot (dy + 1)
\end{aligned}
\right)
\end{equation}

\subsection{Object Direction Classification}\label{sub:directionclassification}

Determining the driving direction of vehicles in images can be achieved through various methods using AI-annotated objects. Traditional approaches often divide the image into zones and classify movement based on the traversed zones. For instance, \cite{Song2019VisionbasedVD} introduces a method that segments the perceived image with horizontal lines. However, such methods typically require multiple frames to evaluate movement, which is unsuitable for our scenario due to minimising processing time.

We propose a three component detection methods \(D_{bb}\), \(D_{eb}\), and \(D_{mod}\) for application to a single frame (image):

\begin{enumerate}
    \item \textit{Static Bounding Box (\(D_{bb}\))}: This method assumes a static view of the scenery, allowing service operators to define bounding boxes that align with street lanes to classify vehicle movements.
    \item \textit{Emerging from the Bottom (\(D_{eb}\))}: This method relies on the observation that objects moving away from the camera first appear at the bottom of the image. As these objects travel further, their detection bounding box increases compared to distant objects.
    \item \textit{Custom Model using Front/Rear View (\(D_{mod}\))}: This method classifies vehicles based on their orientation, using a convolutional neural network (CNN) to distinguish between front and rear views. It requires two additional classification types to infer the vehicle's direction.
\end{enumerate}

Each detection method assigns a value of \(+1\) if the vehicle is heading toward the sensor and \(-1\) if it is moving away. While none of these methods individually provides complete accuracy, we enhance the implementation by combining all three methods as follows:
\vspace{-3mm}
\begin{equation} \label{eq:detectionMethod}
D = D_{bb} + D_{eb} + D_{mod}
\end{equation}

If \(D \geq 1\), the vehicle is determined to be heading toward the sensor; otherwise, it moves away.
\section{Scenario}\label{sec:Scenario}
\begin{figure}[t]
    \centering
    \includegraphics[width=0.48\textwidth]{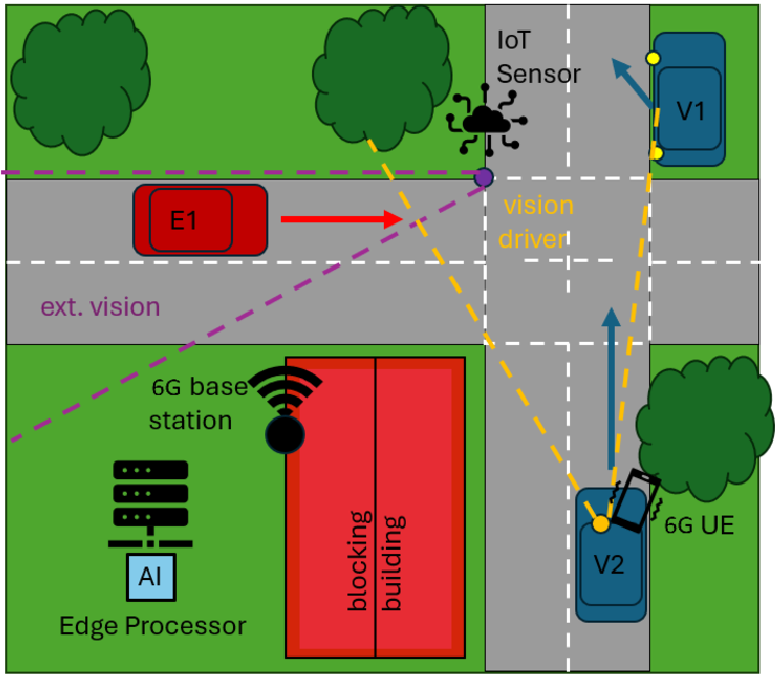} 
    \caption{Traffic scenario of extended view using 6G and AI enabling technology.}
    \label{fig:traffic_scenario}
    \vspace{-6mm}
\end{figure}

The defined scenario integrates the concept of an open sensor network with geographically corresponding users\cite{kitanov22}. The framework is designed for a vehicular context, as shown in Figure \ref{fig:traffic_scenario}. Vehicles \(V_1\) and \(V_2\) operate within this scenario, each equipped with 6G User Equipment (UEs). A 6G base station strategically positions itself along the roadway to ensure low-latency coverage, while a nearby Processing module (Edge) enhances thread classification.

The Processing module acquires raw data from IoT sensor module, serving as the foundation of the classification system. The raw data  includes image data, enabling the detection of critical objects, such as the emergency vehicle (\(E_1\)) depicted in the scenario in Figure \ref{fig:traffic_scenario}.

\subsection{Threat Applicability}\label{sub:applicability}
Not all roadway hazards carry the same significance for every traffic participant. As depicted in Figure \ref{fig:traffic_scenario}, vehicle \(E_1\), approaching from the left, directly influences the risk evaluation for \(V_2\). However, assuming \(E_1\) maintains its current trajectory, it does not affect the risk calculation for \(V_1\). The assessment framework for risk applicability is divided into two distinct categories:

\paragraph{Proximity Events}\label{par:proximity}
This category includes events that impact users unidirectionally within a defined cell. Examples are collisions of other vehicles or fires. While the user may not be directly involved or visually aware of the event, these occurrences can disrupt traffic patterns and influence driver behaviour on the roadway.
Proximity events are consistently treated as warnings to users, which reduces the demand for stringent real-time processing.

\paragraph{Intersection Events}\label{par:intersectionEvent}
Building on the concept of proximity events, intersection events involve incidents aligned with or intersecting the travel vectors of the user. These scenarios are particularly significant at road intersections or multi-use paths where traffic flows converge. Addressing such events requires dynamic risk assessment model to predict and mitigate potential hazards effectively, given the complexity and variability of these environments.

\subsection{Threat Classification}\label{sub:classification}
Based on the applicability of risk, we define the following threat levels:

\paragraph{Alarm}\label{par:alarm}
An \texttt{Alarm} is triggered only when the trajectory vectors of User \(U\) and Sensor \(S\) intersect within the boundaries of the user's cell. This requires the user and the sensor to reside in the same cell \(C\). The definition of the condition is as follows:
\begin{equation} \label{eq:intra}
\mathtt{intra}(U,S) = \begin{cases} 
I \in C(U) \cap C(U) = C(S) &\texttt{true}, \\
I \notin C(U) &\texttt{false}, \\
\nexists I &\texttt{false}.
\end{cases}
\end{equation}

An alarm does activated if the intersection \(I\) occurs in neighbouring cells, outside the grid \(G\), or if the trajectory vectors of \(U\) and \(S\) do not intersect.

\paragraph{Warning 1}\label{par:warning1}
A \texttt{Warning 1} is issued when the trajectory vectors of User \(U\) and Sensor \(S\) intersect within the boundaries of the user's cell, even if the sensor resides in a neighbouring cell. The definition of the condition is as follows:
\begin{equation} \label{eq:inter}
\mathtt{inter}(U,S) = \begin{cases} 
I \in C(U) \cap C(U) \neq C(S) &\texttt{true}, \\
I \notin C(U) &\texttt{false}, \\
\nexists I &\texttt{false}.
\end{cases}
\end{equation}
A \texttt{Warning 1} will not activate if the intersection \(I\) lies within non-adjacent cells, goes beyond the grid boundaries, or if the trajectory vectors do not intersect.

\paragraph{Warning 2}\label{par:warning2}
A \texttt{Warning 2} is raised only when User $U$ and Sensor $S$ reside within the boundaries of the user's cell and if the Sensor event $E$ occurs.
  \begin{equation} \label{eq:coexist}
  \mathtt{coexist}(U,S) = \begin{cases} 
  C(U) = C(S) \cap E(S)  &\texttt{true} ;\\
  C(U) = C(S) &\texttt{false} \\
  \end{cases}
  \end{equation} 

A \texttt{Warning 2} will not be triggered if $U$ and $S$ are in the same cell. Moreover, any intersections do not influence this warning.

\section{Requirement Design for 6G Network}\label{sub:requirements}
This work establishes the networking criteria, focusing on latency, that 6G networks must satisfy to support the proposed traffic safety framework. First, we define the term "real-time" within the context of traffic safety applications. Next, we calculate the computing time for the sensor processing and user module. Finally, we determine the remaining time available for networking delay, represented by $T_{eval}$, $T_{exe}$, and $T_{act}$.

\subsection{Real-Time for Traffic Safety}\label{sub:realtime}
Section \ref{sub:realtime} examines the factors influencing road safety and timely classification, such as vehicle type, road condition, time of day, and speed. The reviewed literature in Section \ref{sec:relatedWork} identifies a suitable time range of 150 to \SI{200}{\milli\second} for the total processing time $T_{tot}$ in our use case.

We apply this definition to our scenario and assess its applicability. Based on data from the sensor module (Section \ref{par:sensorModule}), we estimate that emergency vehicles can be detected within a visual range of \SI{400}{\meter}, considering increased elevation. The speed of an emergency vehicle heavily depends on the traffic situation. Gama \cite{gama18} estimates the vehicle speed at around \SI{20}{\meter\per\second}, although this value can vary.

In our analysis, we assume equal velocities for both the emergency and user vehicles: \( v_E, v_U = 20 \, \text{m/s} \). Kolla \cite{kolla20} reports a braking distance of \SI{27}{\meter} for a modern vehicle (Tesla Model S 90) under a deceleration rate of \( a_b = \SI{8.53}{\meter\per\second^2} \), allowing the vehicle to come to a complete stop from \( v_E \) in approximately \SI{2.83}{\second}.

Using the maximum allowable latency, as defined in Section \ref{par:realtimeDefinition}, we calculate impact velocities of \SI{4.12}{\kilo\meter\per\hour} and \SI{5.49}{\kilo\meter\per\hour} as depicted in Table \ref{tab:impactvelocity}.

\begin{table}[h!]
\centering
\begin{tabular}{|c|c|c|}
\hline
\textbf{Latency (ms)} & \textbf{Impact Velocity (m/s)} & \textbf{Impact Velocity (km/h)} \\
\hline
150* & 1.1445 & 4.1202 \\
200* & 1.526 & 5.4936 \\
250 & 1.9075 & 6.873 \\
300 & 2.289 & 8.2484 \\
350 & 2.6705 & 9.614 \\
400 & 3.052 & 11.0352 \\
\hline
\end{tabular}

\caption{Impact Velocity with Different Latencies (* latency defined by use case)}
\label{tab:impactvelocity}
\end{table}

This finding is significant, as Cormier's \cite{Cormier18} evaluations show an acceleration of \textbf{5 g} at a velocity difference of \SI{5}{\kilo\meter\per\hour} and \textbf{12 g} at a velocity difference of \SI{10}{\kilo\meter\per\hour} at the head of the driver. This data validates our initial estimation based on the literature review and defines our maximum tolerable delay from occurrence to user notification as \( T_{max} = 150 \, \text{ms} \).

\subsection{Computing Time}\label{sub:computingtime}
The computing time is the sum of the individual delays on the three main modules while computing/processing our usecase as depicted in Figure \ref{fig:referenceImages}.
\paragraph{Sensor Module - $T_S$}\label{par:compSensorMod} is defined by two steps, first the image acquisition from the CMOS which usually takes \SI{2}{\milli\second} ($T_{cmos}$) \cite{heo2018super} and the image encoding to MJPEG which is expected to contribute \SI{10}{\milli\second} to \SI{20}{\milli\second} ($T_{enc})$ according to Song \cite{Song13}. MJPEG is still widely use in areas where low latency video encoding is required \cite{Pawłowski24}. This lets us define $T_S$ with \SI{22}{\milli\second}.
\paragraph{Processing Module - $T_P$}\label{par:compSensorMod} is composed of three steps. First the extraction of the image from the input stream, this is very simple using MJPEG showing less than \SI{1}{\milli\second} ($T_{dec}$). Next part is the processing time the CNN demands to detect vehicles $T_{AI}$, this is in depth explored in Section \ref{sec:Results}
and finally the method to perform the thread assessment as defined in \ref{sub:classification} which is also empirical evaluated in Section \ref{sub:resultThreadClass} and designated as $T_{TC}$
\paragraph{Consumer Module - $T_C$}\label{par:compSensorMod}
The Consumer module receives an event notification from the Processing module using a pre-established socket connection. In our case, we refer to measurements Maassen \cite{Maassen99} conducted, which show \SI{1200}{\micro\second}. We define that only an acoustic alarm will be triggered and assign $T_C$ a value \SI{1}{\milli\second}. In this evaluation, we do not conduct further on how long it would take until the driver sets an action since we only focus on the technical parameters here. However, we want to reference the work of Lewis \cite{lewis2018validation} and Wiese \cite{wiese2004auditory}, which did extensive work on this aspect.

\subsection{6G Networking Latency}\label{sub:6gnetworkinglatency}
The 6G networking latency is a derived metric from the total processing time $T_{tot}$, which we discussed in Section \ref{sub:realtime}, and is reduced by all other delays observed during the execution of the use case:
\vspace{-3mm}
\begin{equation} \label{eq:6gnetcomb}
(T_{eval} + T_{exe}) = T_{tot} - ( T_{S} + T_P + T_C). 
\end{equation} 

Since $(T_{eval} + T_{exe})$ represents the combined latency from the Sensor module to the Processing module and from the Processing module to the user module, we need to distribute the two terms. Assuming that the Sensor and Consumer modules reside in the same cell, we split the latency equally. This leads to the following definitions for $T_{eval}$ and $T_{exe}$:

\begin{equation} \label{eq:6gnetseperated}
T_{eval} = T_{exe} = \frac{T_{tot} - ( T_{S} + T_P + T_C)}{2}. 
\end{equation}  

 \section{Framework}\label{sec:framework}
 We introduce the framework design and its logical components in the implementation.





\subsection{Implementation: Emergency Vehicle Warning System}\label{sub:emergencyvehwarning}
Based on the scenario described in Section \ref{sec:Scenario}, the implemented solution detects emergency vehicles and provides a \textbf{warning} signal to all users within a \SI{2}{\kilo\meter} radius. It also sends an \textbf{alarm} signal to all vehicles that will intersect with the user's path. In Figure \ref{fig:traffic_scenario}, the driver of vehicle V2 approaches an intersection, but a building on the left obstructs their view as an emergency vehicle speeds towards them. The IoT sensors, equipped with a camera, monitor the orthogonal direction in which vehicle V2 is travelling. The IoT sensor transmits the camera feed to the Edge Processor, which must evaluate whether a \textit{Warning} and/or \textit{Alarm} action is necessary. In our scenario, vehicles V2 and V1 receive a warning and information about their trajectories. The user equipment (UE) itself makes the decision to trigger an alarm.Therefore, vehicle V2 will trigger an alarm, as it could collide with the emergency vehicle E1. Vehicle V2 will only display a warning, as no imminent threat of collision with E1 exists.

\paragraph{Sensor Module}\label{par:sensorModule} is deployed on a Raspberry Pi v4, gathering full-HD image data. It uses MJPEG for low-latency video encoding. The sensor streams the data over a previously established TCP connection to the processing module. The connection is initiated by the processing module, followed by a registration request, which includes the following parameters:

\begin{itemize}
    \item \textit{deviceID}: A unique identifier assigned to the sensor module.
    \item \textit{SocketIn}: The socket through which the sensor expects connection requests to start the video stream.
    \item \textit{Trajectory}: The direction in which the sensor detects incoming vehicles. For example, if the camera is oriented north (0 degrees), all incoming vehicles have a trajectory of 180 degrees. If the camera faces west, the Trajectory would be 90 degrees.
    \item \textit{Zone}: Defines the zone covered by the sensor. We segment the applicable area of the sensor network into a grid of 1 km² square cells. The cells are designated horizontally with letters and vertically with numbers. For example, we could designate cells as B3, G17, etc.
\end{itemize}

\begin{figure}[ht]
    \centering
    \begin{subfigure}[b]{0.24\textwidth}\includegraphics[scale=0.11]{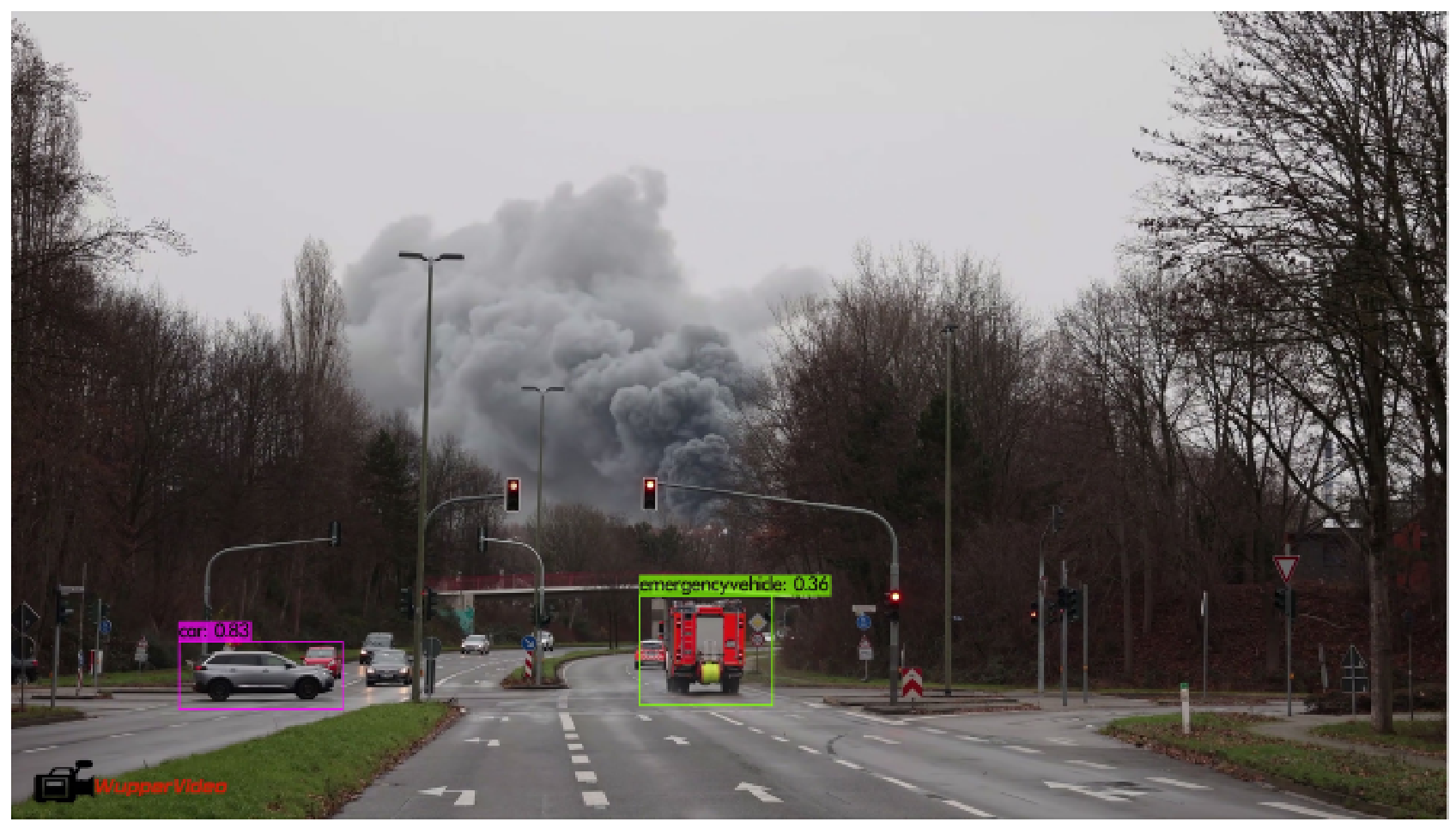}\caption{Image 1.png}\label{fig:img1}\end{subfigure}
    \begin{subfigure}[b]{0.24\textwidth}\includegraphics[scale=0.11]{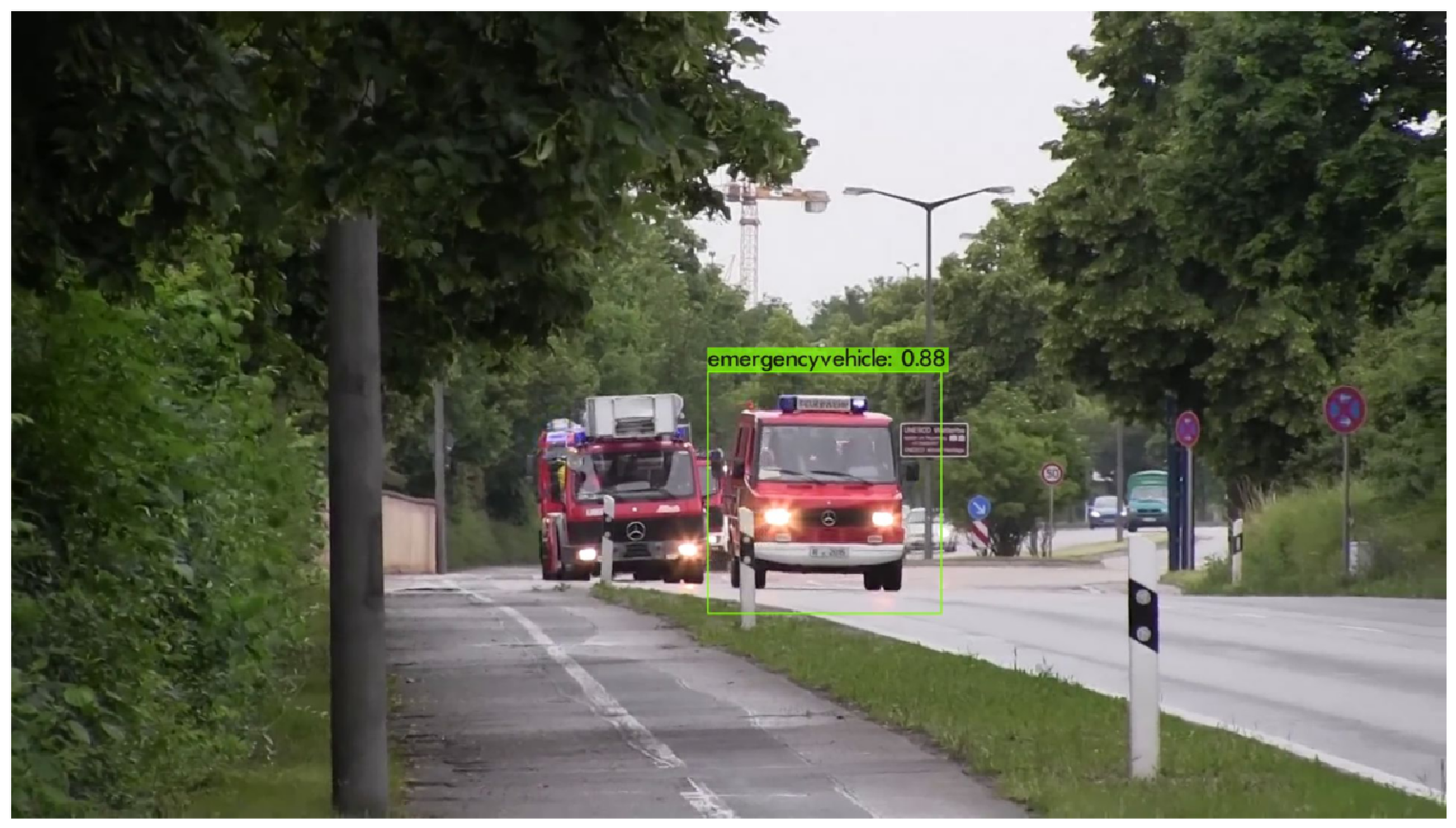}\caption{Image 9.png}\label{fig:img9}\end{subfigure}
    \begin{subfigure}[b]{0.24\textwidth}\includegraphics[scale=0.11]{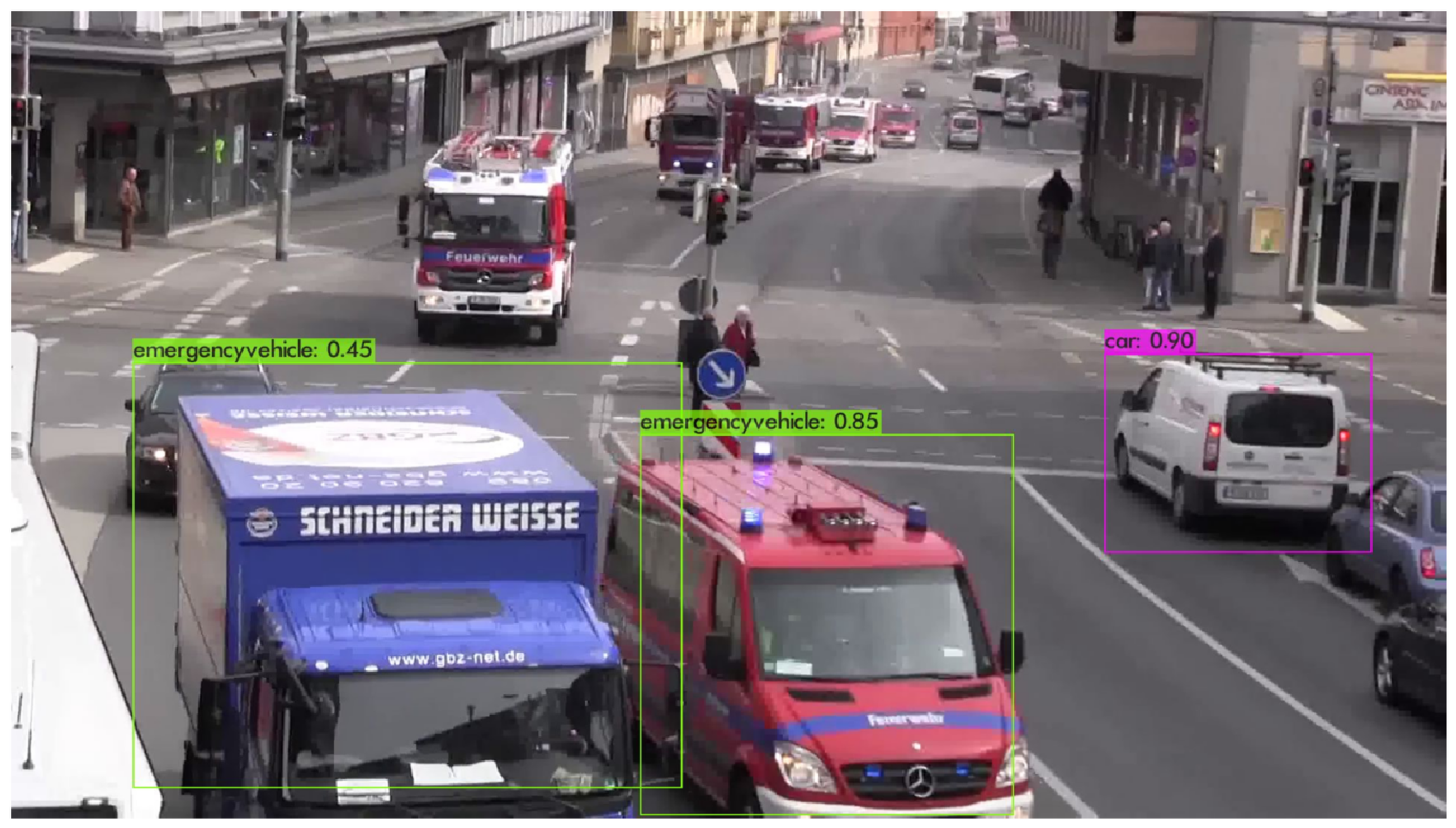}\caption{Image 10.png}\label{fig:img10}\end{subfigure}
    \begin{subfigure}[b]{0.24\textwidth}\includegraphics[scale=0.11]{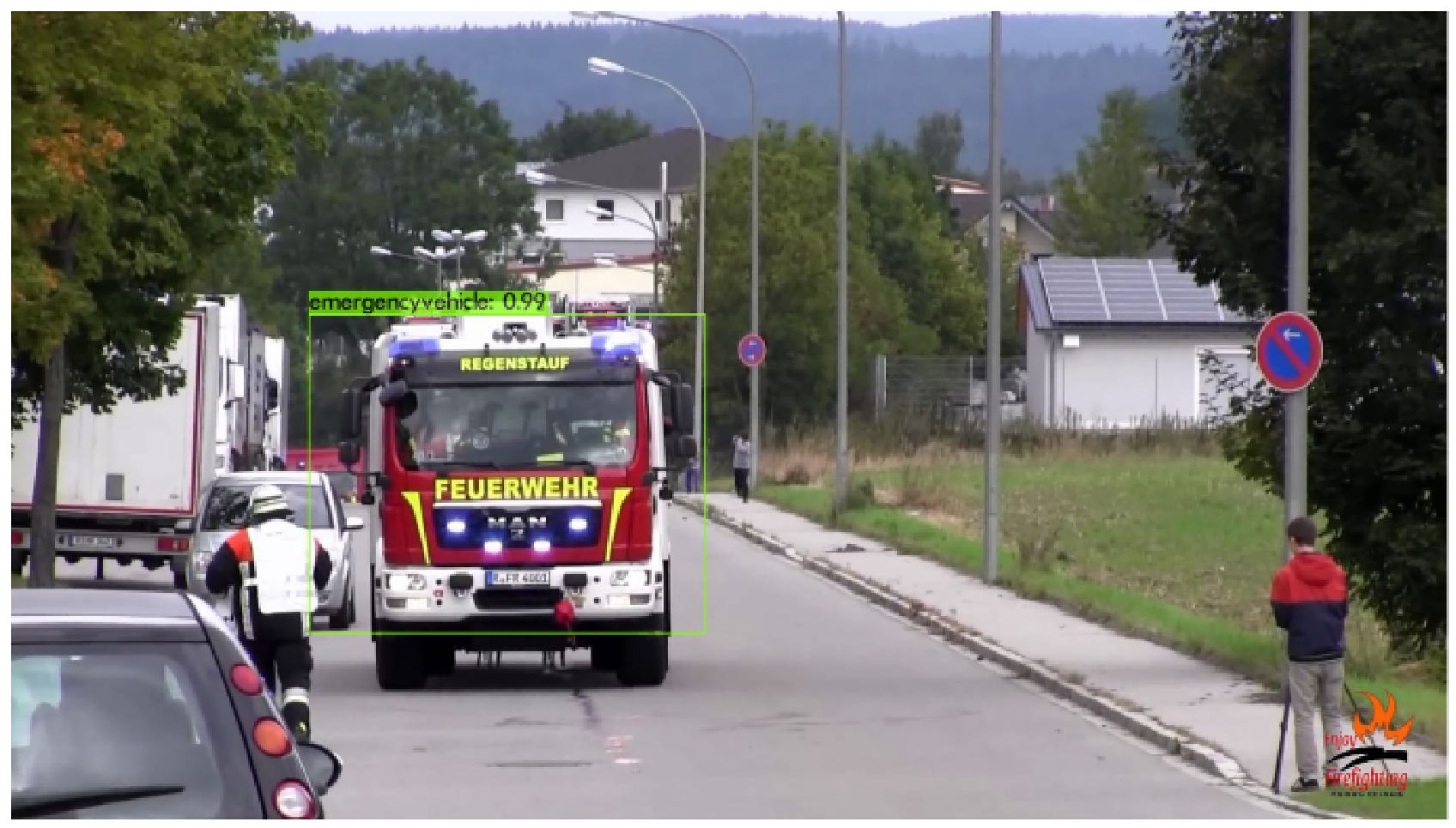}\caption{Image 14.png}\label{fig:img14}\end{subfigure}
    \begin{subfigure}[b]{0.24\textwidth}\includegraphics[scale=0.11]{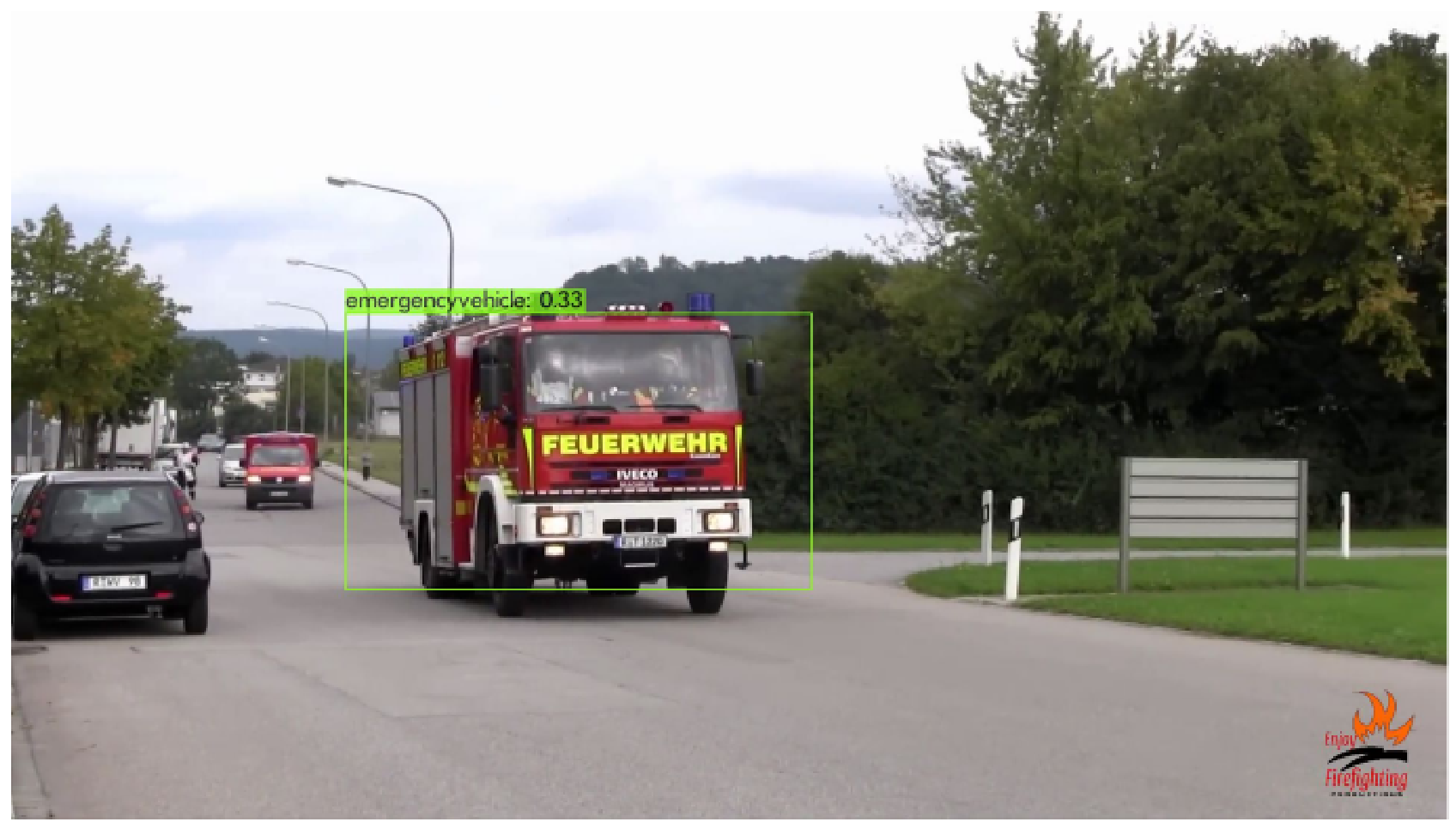}\caption{Image 16.png}\label{fig:img16}\end{subfigure}
    \begin{subfigure}[b]{0.24\textwidth}\includegraphics[scale=0.11]{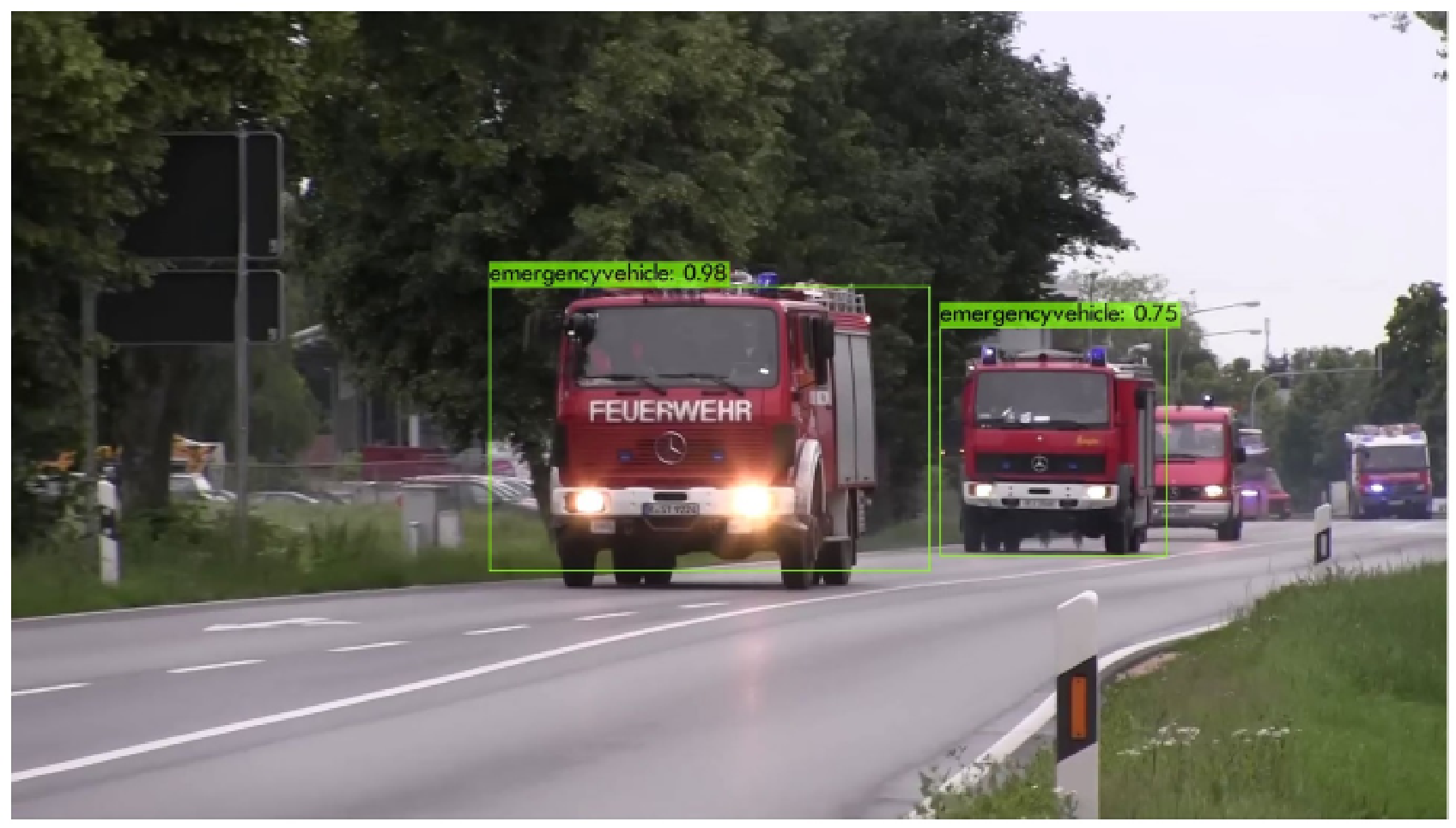}\caption{Image 18.png}\label{fig:img18}
    \end{subfigure} 
    \caption{Selection of reference images used. The labelled boxes indicate vehicle classification.}
    \label{fig:referenceImages}
    \vspace{-1mm}
\end{figure}

A set of reference data acquired by the sensor module is shown in Figure \ref{fig:referenceImages}.

\paragraph{User Module} is deployed on a Xiaomi Redmi Note 13 running Android 13. The user module is implemented as a native application and is registered with the processing module using MESDD, a service discovery scheme that utilizes geofences and DNS \cite{mesdd23}. In the first stage, the User module registers itself with the Processing module and establishes a TCP connection. This connection allows the Processing module to forward the trajectory and location of the sensors, provided that both the User and Sensor modules reside within the same grid. Thread classification can then be applied as described in Section \ref{sub:classification}. 

Classifying the User module thread demands no forward of privacy-sensitive information, despite its exact position and trajectory, to the Processing module. 

To associate with the correct sensor data, the user module must provide the following details and update them if the current cell changes:

\begin{itemize}
    \item \textit{deviceID}: A unique identifier generated by the user module (no personal data is required).
    \item \textit{SocketIn}: The socket on which the user module expects incoming connection requests.
    \item \textit{Zone}: Defines the user's location. A transition between zones must trigger an update of the user's information.
\end{itemize}

\paragraph{Processing Module} is the core component of our implementation, linking the user with the sensor for determining thread applicability. To support this decision-making process, a CNN is employed based on the image data provided by the sensor. The image classification relies on a customized CNN, which uses YoloV4 as the detector. The bounding boxes assign specific vehicle classes to the detection results. We optimized our model to detect emergency vehicles from various angles as delineated in our dataset\footnote{\url{https://github.com/kurthorvath/EmergencyVehicleYoloY4}}.

The Processing module determines the sensor's basic vision trajectory during the sensor registration process. However, if a vehicle moves towards or away from the sensor, this must be identified during the classification process, as detailed in Section \ref{sub:directionclassification}.


 \section{Results}\label{sec:Results}
 The evaluation of novel and unique components such as vehicle classification focuses distinctly on their contributions to latency. We assess the performance of emergency vehicle detection by comparing results across the test set. The analysis of the thread classification process follows and examines its impact on overall delay.

\subsection{AI Inter-Frame Stability}\label{sub:interframehist}
\begin{figure}[t]
\vspace{-5mm}
   \centering
    \includegraphics[width=0.45\textwidth]{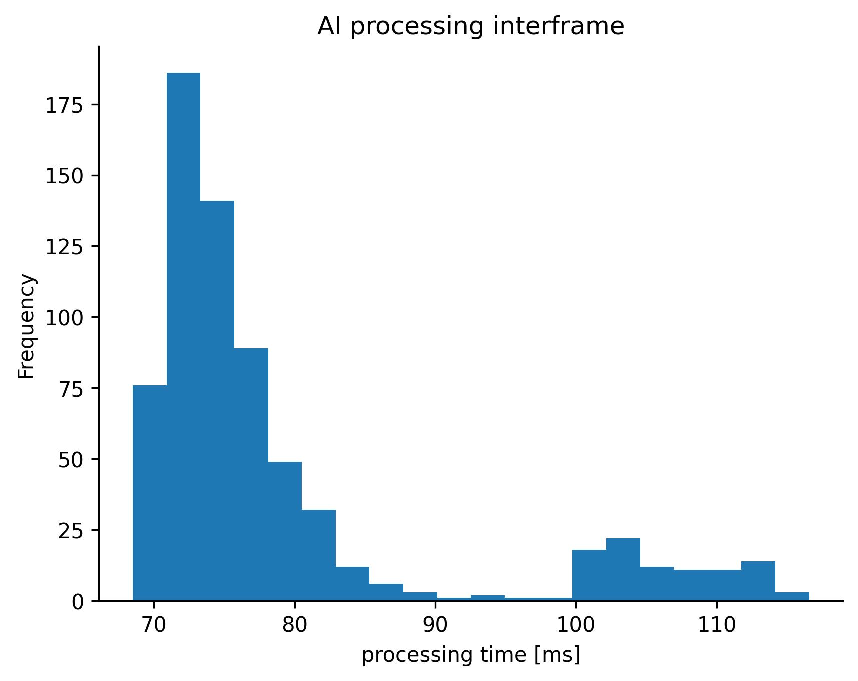} 
\caption{AI performance applying reference data}
    \label{fig:interframe}%
\end{figure}

Figure \ref{fig:interframe} illustrates the distribution of processing times for all reference images. Most images are processed within a time window of 70 to \SI{80}{\milli\second}, with the core processing time ranging from 72 to \SI{75}{\milli\second}. This consistent pattern indicates that the AI model reliably operates within a predictable timeframe, effectively managing most images to identify potential threats to traffic flow within a smart city infrastructure.

However, the distribution includes outliers where processing times exceed \SI{100}{\milli\second}, with the maximum recorded time reaching \SI{105.52}{\milli\second}. These anomalies, represented by the right-skewed tail of the distribution, suggest that specific images pose elevated computational challenges, potentially due to increased complexity or other system demanding on the processing node.

\begin{table}
  \centering
\parbox{.45\linewidth}{
\caption{AI computing time [ms]}
    \label{tab:AIprocessing}
    \begin{tabular}{|l|l|}
    \hline
        \textbf{Statistic} & \textbf{$T_{AI}$ (ms)} \\ \hline
        Mean & 79.02 \\ 
        Median & 74.49 \\ 
        Standard Deviation & 11.53 \\ 
        \hline
    \end{tabular}
}
\hfill
\parbox{.45\linewidth}{
    \caption{thread classification time [ms]}
    \label{tab:Threadclassification}
    \begin{tabular}{|l|l|}
    \hline
        \textbf{Statistic} & \textbf{$T_{TC}$ (ms)} \\ \hline
        Mean & 0.95 \\ 
        Median & 0.90 \\ 
        Standard Deviation & 0.21 \\ 
        \hline
    \end{tabular}
}
\vspace{-2mm}
\end{table}

\subsection{Performance of Thread Classification}\label{sub:resultThreadClass}
The classification process identifies potential intersections between an emergency vehicle and a user, as detailed in Section \ref{par:intersectionEvent}. To assess the performance of this process, we implemented a reference application for the intersection calculation and executed it 1000 times.

The results, summarized in Table \ref{tab:Threadclassification}, indicate that both the mean and median processing times are below \SI{1}{\milli\second}. The mathematical simplicity of the intersection calculation ensures rapid and accurate processing while minimizing computational overhead, contributing to its high efficiency.

\begin{figure}[t]
   \centering
    \includegraphics[width=0.45\textwidth]{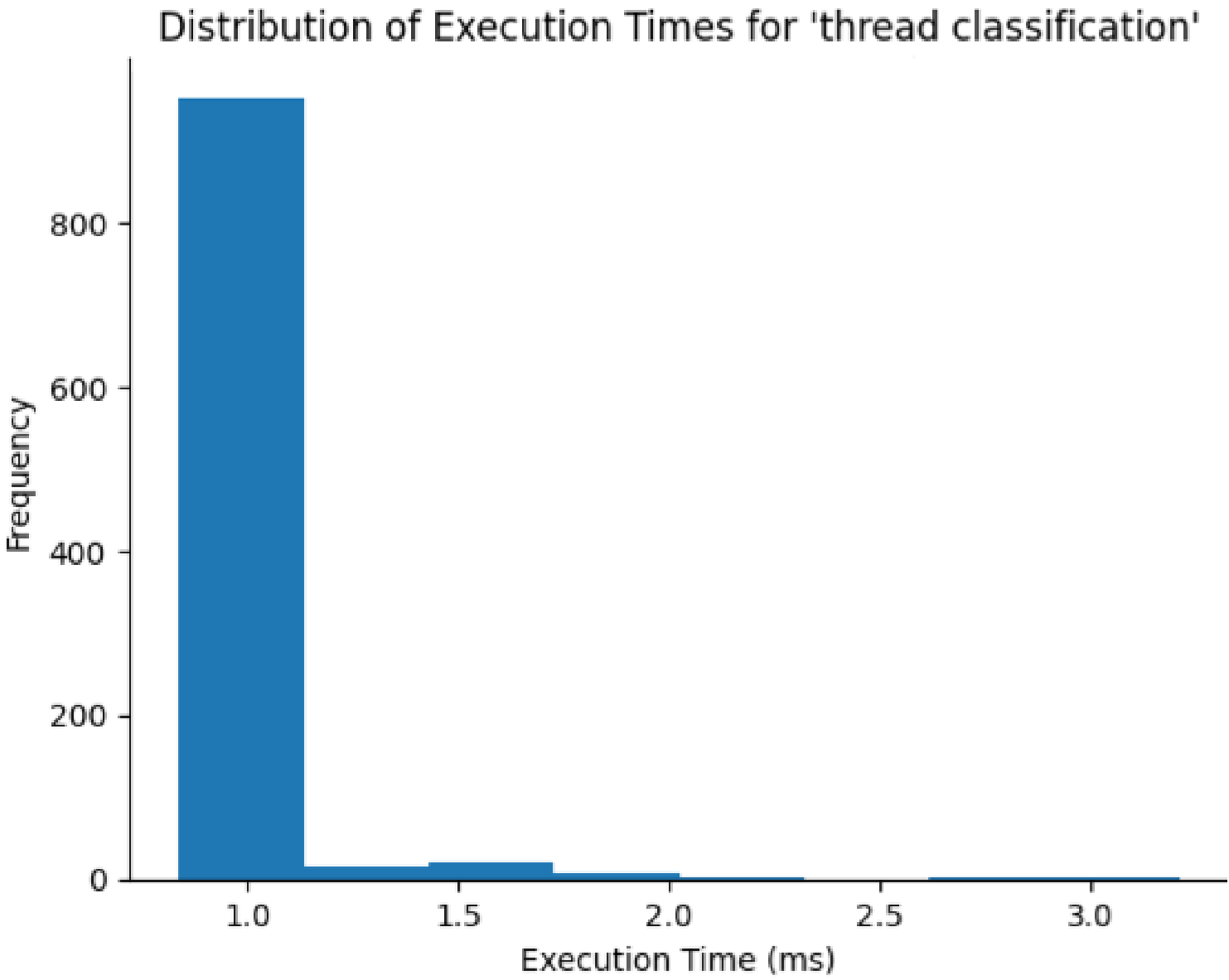} 
\caption{Performance thread classification}
    \label{fig:threadclass}%
    \vspace{-6mm}
\end{figure}

\section{Conclusion}\label{sec:Conclusion}
In this work, we established a model to describe the temporal components of a 6G-enabled framework designed for traffic safety, utilizing AI to extend users' visual range to enhance their driving safety. Equation \ref{eq:6gnetseperated} defines the tolerable network latency in contrast to the real-time requirements ($T_{tot}$) outlined in Section \ref{sub:realtime}. All technical contributions unrelated to networking have been analyzed in detail and are summarized as follows:

\begin{equation} \label{eq:Tsensor}
T_{S} = T_{cmos} + T_{enc} = 22 \, \text{ms}
\end{equation} 
\vspace{-2mm}
\begin{equation} \label{eq:Tprocessing}
T_{P} = T_{dec} + T_{AI} + T_{TC} = 76.39 \, \text{ms}
\end{equation} 

The Consumer module ($T_C$) processes events within \SI{1.2}{\milli\second}. 

Using these values, we can now evaluate Equation \ref{eq:6gnetseperated}, determining the tolerable latency for the networking components in the proposed framework.

\begin{equation} \label{eq:Result}
T_{eval} = T_{exe} = 25.20 ms
\end{equation} 

In respect to Popovski \cite{popovski2022perspective} and Kitanov \cite{kitanov2021186}, we expect round trip latency of less \SI{1}{\milli\second} which suffice our demand of \SI{25.2}{\milli\second} but as also Popovski states in the same work it is expected that 5G operates also bellow \SI{5}{\milli\second} where current real-world studies show that results exceed these values by far \cite{Rischke21}.

Infrastructure providers and operators must not only provide low latency networking performance on campus networks or experimental setups but also be integrated into the existing infrastructure so use cases like the line of sight extension we propose is feasible, considering that lower latency beyond the threshold we define still improves safety.

\section*{Acknowledgement}
This work received funding from the Austrian Research Promotion Agency (FFG) (grant 888098 ``K{\"a}rntner Fog'' and grant 909989 ``AIM AT Stiftungsprofessur für Edge AI'') and the OeAD for collaboration of the University of Klagenfurt (Austria) and Mother Teresa University (Republic of North Macedonia)

\bibliographystyle{IEEEtran}
\bibliography{mybib}

\end{document}